\documentclass[aps,prl,showpacs,amsmath,twocolumn,amssymb,superscriptaddress,letterpaper]{revtex4}

\usepackage{graphicx,color}
\usepackage{amssymb}   
\usepackage{amsmath}
\usepackage{epstopdf}
\usepackage{natbib}
\usepackage{hyperref}
\usepackage{bm}

\begin{document}

\preprint{APS/123-QED}

\title{Unveiling nontrivial fusion rule of Majorana zero mode using a fermionic mode}

\author{Yu Zhang}
\affiliation{School of Physics, MOE Key Laboratory for Non-equilibrium Synthesis and Modulation of Condensed Matter, Xi’an Jiaotong University, Xi’an 710049, China}

\author{Xiaoyu Zhu}
\affiliation{School of Physics, MOE Key Laboratory for Non-equilibrium Synthesis and Modulation of Condensed Matter, Xi’an Jiaotong University, Xi’an 710049, China}

\author{Chunhui Li}
\affiliation{School of Physics, MOE Key Laboratory for Non-equilibrium Synthesis and Modulation of Condensed Matter, Xi’an Jiaotong University, Xi’an 710049, China}

\author{Juntao Song}
\affiliation{Department of Physics and Hebei Advanced Thin Film Laboratory, Hebei Normal University, Shijiazhuang 050024, China}

\author{Jie Liu}
\thanks{Corresponding author: jieliuphy@xjtu.edu.cn}
\affiliation{School of Physics, MOE Key Laboratory for Non-equilibrium Synthesis and Modulation of Condensed Matter, Xi’an Jiaotong University, Xi’an 710049, China}
\affiliation{Hefei National Laboratory, Hefei 230088, China}

\author{X. C. Xie}
\affiliation{International Center for Quantum Materials, School of Physics, Peking University, Beijing 100871, China}
\affiliation{Institute for Nanoelectronic Devices and Quantum Computing, Fudan University, Shanghai 200438, China}
\affiliation{Hefei National Laboratory, Hefei 230088, China}

\begin{abstract}
  Fusing Majorana zero modes leads to multiple outcomes, a property being unique to non-Abelian anyons. Successful demonstration of this nontrivial fusion rule
 would be a hallmark for the development of topological quantum computation.
  Here we show that this can be done by simply attaching a fermionic mode to a single Majorana zero mode. Through modulation of the energy level of this fermionic mode as well as its coupling with the Majorana mode in different sequences,  we show that a zero or integer charge pumping can be realized when different fusion loops are chosen. Such fusion loops are intimately related with the nontrivial fusion rule of Majorana modes and are solely determined by the crossings at zero energy in the parameter space. Finally we demonstrate our proposal in a nanowire-based topological superconductor coupled to a quantum dot. We show that the charge pumping is robust for MZMs in the real system irrespective of the initial condition of FM state, contrary to the case for trivial Andreev bound states. This provides a feasible way to distinguish Majorana modes from trivial Andreev bound states.
\end{abstract}

\pacs{74.45.+c, 74.20.Mn, 74.78.-w}

\maketitle



\textit{Introduction.}  It has been a long-cherished dream of condensed matter physicists to find Majorana zero mode (MZM) for its potential application in fault-tolerant topological quantum computations \cite{kitaev, nayak}.
 Despite overwhelming evidences in favor of MZMs being reported over the last decade \cite{Fu, sau, fujimoto, sato, alicea2, lut, 2DEG1, 2DEG2, kou, deng, das1, hao1, Marcus, deng2, perge, Yaz2, Jia, Fes1, Fes2,PJJ1, PJJ2}, 
 there is still room for other possible interpretations in regards to experimental results of various studies, such as the existence of Andreev bound states (ABS)\cite{Jie1, brouwer, Aguado, ChunXiao1, Moore, Wimmer, Aguado2,Pan,ABSM}. 
 While it's possible for ABS to demonstrate a great deal of similarities to MZMs in transport experiments, the two are fundamentally different particles in terms of the exchange statistics they obey, which could be exploited to discriminate them \cite{Ivanov, alicea3, NQP, Jienew}.

In contrast to ABS which are ordinary fermions, MZMs follow non-Abelian statistics. This non-Abelian property reveals itself in two distinct but closely related aspects. First, in a system of MZMs, their quantum state may experience a nontrivial rotation when two of them are exchanged, unlike ABS or other Abelian anyons where the state only acquires a global phase. More importantly, two consecutive rotations in differing sequences may incur totally different final states, hence the name \emph{non-Abelian}. Such rotations, or braiding operations, constitute the basic logic gates of topological quantum computations, with various theoretical studies being devoted to in the last decades\cite{TQC2, TQC3, TQC4,MSQ, net1, net2, Yu1,Sato,Roy, QD1}. 
Besides the braiding, fusion is another equally important non-abelian property of MZMs\cite{nayak}. For MZMs, two of them can fuse into a particle of trivial type $I$ or a fermion $\Psi$, with the fusion rule formally expressed as $\gamma\times \gamma = I+\Psi$. Since the outcome is multiple, the final fusion results would rely on the fusion order of MZMs. In principle, since fusion is actually a process that couple two MZMs with finite energy, it should be easier to realize in experiment compared with braiding. Several pioneer work have suggested that fusion would induce nontrivial charge transport \cite{fusion1,fusion2,fusion3}. However, the protocols are still complicated for further investigation.

\begin{figure}
\centering
\includegraphics[width=3.25in]{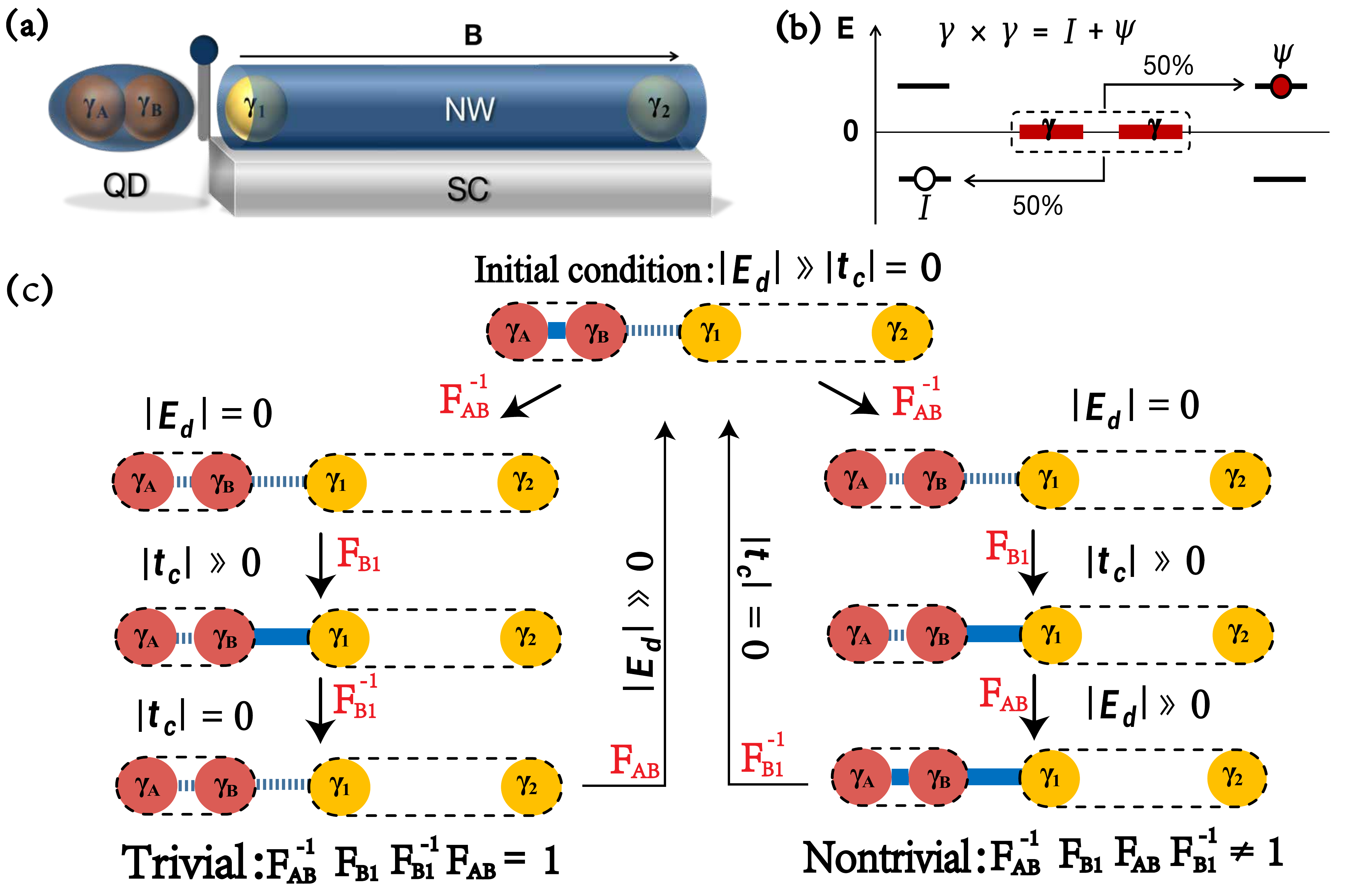}
\caption{
(a) A coupled quantum dot (QD) nanowire (NW) system that could demonstrate our fusion protocol. Quantum dot provides the fermionic mode (equivalent to two MZMs) needed while the unpaired Majorana zero modes are expected to appear at the ends of superconducting nanowire.
(b) The fusion rule of two MZMs. Two fusion outcomes, vacuum state $I$ and unpaired fermionic state $\Psi$, appear with equal probability.
(c) For a typical trivial loop (left panel), we first split $\gamma_A$ and $\gamma_B$ ($F_{AB}^{-1}$), followed by the fusion of $\gamma_B$ and $\gamma_1$ ($F_{B1}$), and then reverse the operation to complete the loop. In a typical nontrivial loop (right panel), two consecutive processes do not commute with each other.}
\label{f1}
\end{figure}

In this paper, we propose to unveil the nontrivial fusion rule of MZMs by simply attaching a fermionic mode (FM) to a single MZM. Such a setup could be possibly realized in multiple systems, for instance, in a nanowire based topological superconductor (TS) that is coupled to a quantum dot \cite{deng2} (see Fig. \ref{f1}(a)), or in a topologically nontrivial vortex with a molecule attached to scanning probe microscopy acting as the FM \cite{Jievortex}. The FM can usually be deemed as two pre-fused MZMs $\gamma_A$ and $\gamma_B$ with the hybridization energy $E_d$. As such, these two MZMs can be effectively split by tuning $E_d$ to zero. 
The fusion and splitting process between $\gamma_B$ and another MZM $\gamma_1$ in the TS can be controlled by tuning the coupling between the FM and the superconductor, as schematically depicted in Fig. \ref{f1}(a). This is because MZM is its own anti-particle, and $\gamma_1$ only couples to half of the FM $\gamma_B$ \cite{Jienew}. Based on these observations, different fusion and splitting processes can be realized through tuning the energy level of FM or adjusting the coupling between MZM and FM. Recent experiments have demonstrated that these parameters can be easily modulated using gate voltage \cite{Pkitaev, Pgate1,Pgate2}, making our fusion protocol feasible for practical systems. Furthermore, these nontrivial fusion processes could induce significant charge transfer events, which can be precisely detected using charge sensing methods \cite{charge}.

\textit{Model-independent Hamiltonian of the setup and the current formula.} To simulate the fusion process of MZMs, we start from a model-independent Hamiltonian:
\begin{equation}
H_s = {2E_d(t)}{d^\dag }d + [{t_c(t)}d - t_c(t)^* {d^\dag }]{\gamma_1} +iE_M{\gamma_1}{\gamma_2},
\label{eq1}
\end{equation}

\noindent where $d$ is the annihilation operator of FM, and $E_d$ is the on-site energy of FM. $\gamma_1$ and $\gamma_2$ are a pair of MZMs coming from TS, and $E_M$ represents the coupling between them. The coupling between FM and $\gamma_1$ is denoted by $t_c=|t_c|e^{i\phi/2}$, in which $\phi$ is the pairing phase of TS \cite{S1}. By decomposing FM operator into superposition of two MZMs, $d = \frac{1}{2}e^{-i\frac{\phi}{2}}(\gamma_A+i\gamma_B)$, Hamiltonian in Eq.(\ref{eq1}) can be rewritten in the following simple form:

\begin{equation}
\begin{aligned}
H_M &= i{E_d}\gamma_A\gamma_B +i|t_c|\gamma_B{\gamma_1}+iE_M{\gamma_1}{\gamma_2}.
\end{aligned}
\label{eq2}
\end{equation}

\noindent From Eq.(\ref{eq2}) we note $E_d$, $E_M$ and $t_c$ can be treated as couplings between different MZMs. This suggests that one could simulate fusion process by tuning these parameters from zero to a finite value, and vice veresa.

The charge transfer of FM can be defined as the integration of the time-dependent current from FM to TS, i.e., $\delta N(t) = \int_0^{t}\langle \hat{J}_e(\tau)+\hat{J}_h(\tau) \rangle \mathrm{d}\tau$ \cite{Jie2, Lei}, with
\begin{equation}
 \langle \hat{J}_{e(h)}(t) \rangle = -\frac{i}{2}
 Tr(\langle \psi_{d}^e(t)|H_c^{e(h)}(t)|\psi_{M}^{e(h)}(t) \rangle + h.c.)
 \label{eq4}
 \end{equation}
\noindent Here, $|\psi_{d}^e(t)\rangle$ is the electron part wavefunction of FM at time $t$, and $|\psi_{M}^{e(h)}(t) \rangle$ is the electron (hole) part wavefunction of MZMs at time $t$. $H_c^e(t) = \frac{t_c(t)}{2}d \Psi_M^{\dag}$ and $ H_c^h(t) =  \frac{t_c(t)}{2}d \Psi_M$, representing two different parts of $t_c$ coupling term in Eq.(\ref{eq1}), with fermionic state $\Psi_M=\gamma_1+i\gamma_2$. Equation (\ref{eq4}) suggests that our system supports two charge transfer processes. The first one, $J_e$, involves an electron hopping from FM to MZM or vice versa (termed as TE process). During the process represented by $J_h$, an electron in FM transforms into a hole in MZM (termed as TH process). 
Which of the two processes to occur is determined by fermion parity of the system.

\begin{figure}
\centering
\includegraphics[width=3.25in]{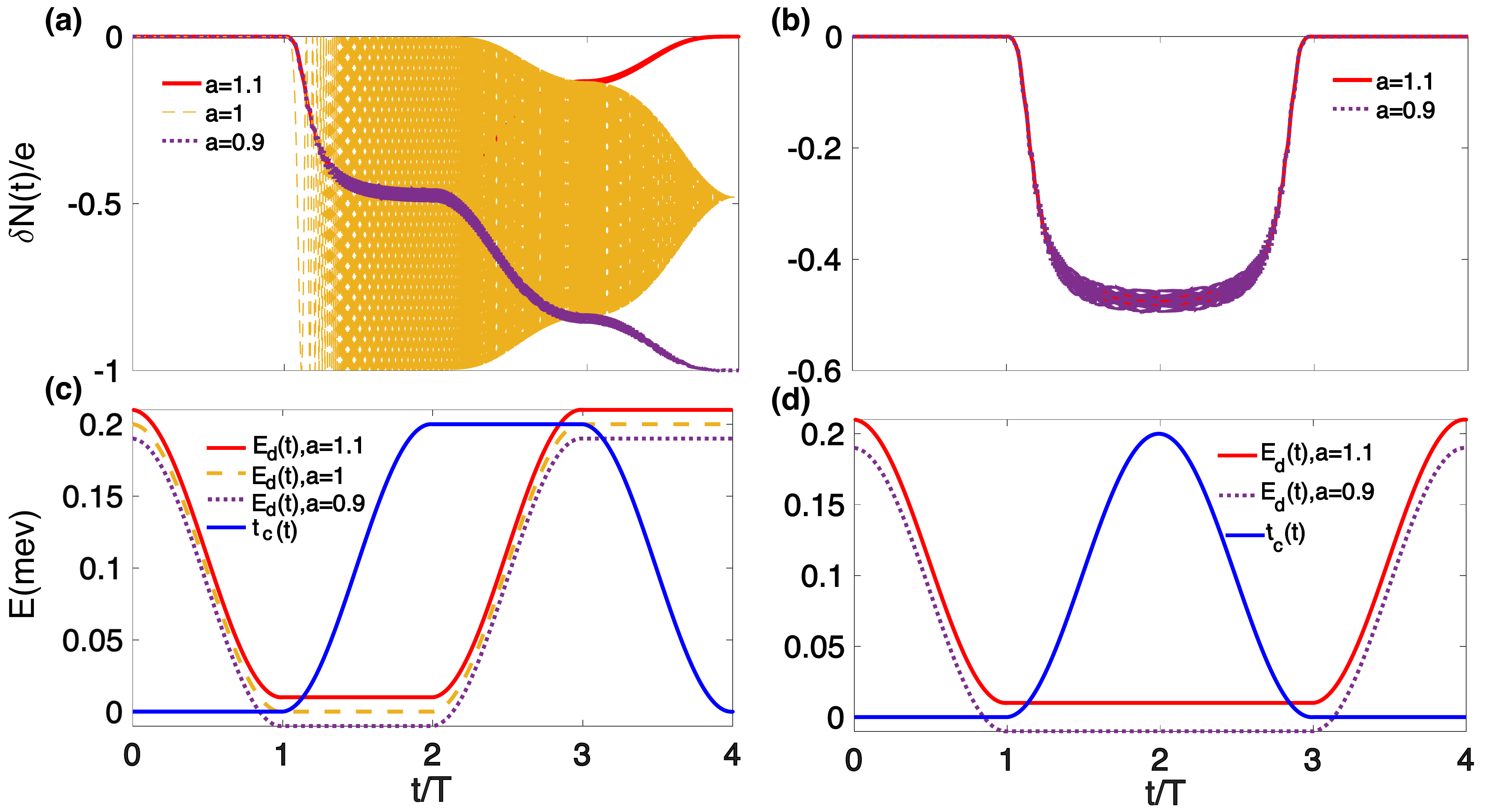}
\caption{
(a)  The charge transfer for the nontrivial loop $F_{AB}^{-1}F_{B1}F_{AB}F_{B1}^{-1}$. Here $F_{AB}^{-1}$ ($F_{AB}$) is controlled by $E_d(t) = E_0[a\pm \cos(\omega t)]$ with $\omega=1/1000 \text{ mev}$, $T = \pi/\omega$, and $E_M=10^{-9} \text{ mev}$. After the splitting process, the energy level of the FM reaches exactly to zero for $a = 1$, stays above zero for $a = 1.1$, and changes sign for $a=0.9$.
(b) The charge transfer for the trivial loop $F_{AB}^{-1}F_{B1}F_{B1}^{-1}F_{AB}$.
(c) and (d) show the evolution of system parameters corresponding to (a) and (b), respectively.}\label{f2}
\end{figure}

\textit{Nontrivial fusion loop and the related charge transfer.}   Due to the nontrivial fusion rule of MZMs, the fusion outcomes would rely on their fusion order. To distinguish between fusion and splitting processes, we denote the fusion of MZM $i$ and MZM $j$ as $F_{ij}$, and the splitting process as $F_{ij}^{-1}$. It is clear that when fusion and splitting of two MZMs occur consecutively, the end result is trivial, i.e.,
$F_{ij}^{-1}F_{ij}=I$. A minimal model illustrating the nontrivial fusion rule requires at least four MZMs. The left loop in Fig. \ref{f1}(c) depicts a prototypical trivial loop for this minimal model since $F_{AB}^{-1}F_{B1}F_{B1}^{-1}F_{AB}=I$. A nontrivial fusion loop, however, requires two consecutive processes to be non-commute, such that the final state cannot return to the initial one after a complete loop. Based on this rule, many nontrivial loops can be designed by manipulating the order of fusion and splitting processes. The right loop in Fig. \ref{f1}(c) illustrates a representative nontrivial loop, denoted by $F_{AB}^{-1}F_{B1}F_{AB}F_{B1}^{-1}$, which is similar to the one introduced in Ref. \onlinecite{fusion1}. However, in this new proposal we need not tune the parameters in the TS system, which may avoid unwanted adverse effects due to the manipulation of TS wires.

 
\begin{figure}
\centering
\includegraphics[width=3.25in]{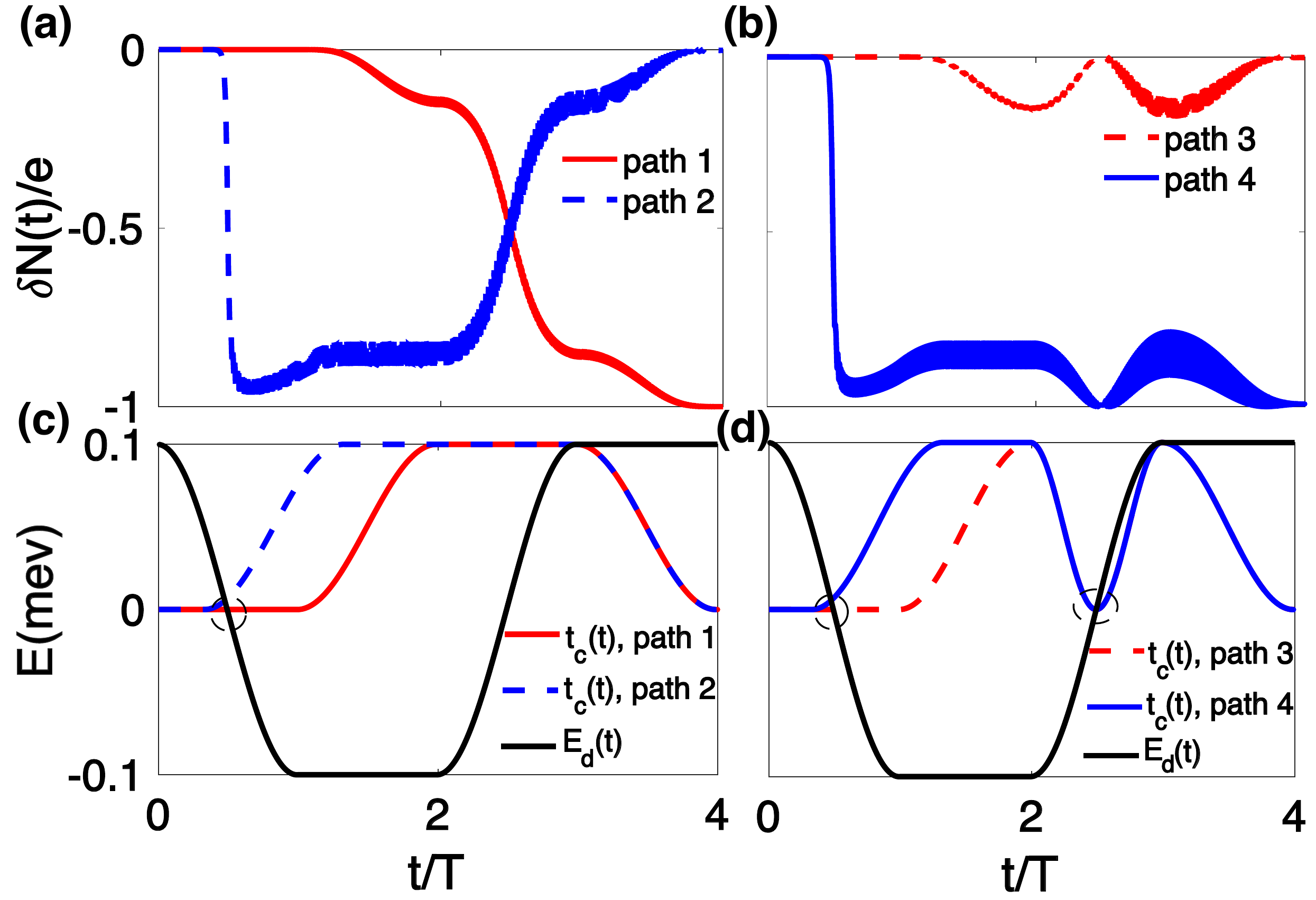}
\caption{The charge pumping would be nontrivial if the parameters $E_d$ and $t_c$ cross at zero energy with odd times.
(a) and (b) show the corresponding charge transfer for paths depicted in (c) and (d). These paths only differ in how $t_c$ evolves in the corresponding fusion loop. Clearly, $E_d$ and $t_c$ cross at zero energy with an odd number of times in path 1 and 2.
}
\label{f3}
\end{figure}

To validate our proposal, we numerically calculate the charge transfer during both the trivial and the nontrivial loops. The splitting of the two MZMs that constitute the FM is controlled by $E_d$, which we assume to vary with time according to $E_d(t) = E_0[a + \cos(\omega t)]$ with $a=1$. This suggests that the FM would initiate with energy $2E_0$ at $\omega t= 0$ and split at zero energy at $\omega t= \pi$. The fusion process is just the reverse of this process with $E_d(t) = E_0[1- \cos(\omega t)]$. Similarly, $t_c(t)=|E_0|[1\pm \cos(\omega t)]$ controls the splitting and fusion process of $\gamma_B$ and $\gamma_1$. At initial time we have $E_d=2E_0>0$ and thus the system is in $|0_{\text{AB}}0_{12} \rangle $ state. We may calculate the charge transfer for the nontrivial loop denoted by $F_{AB}^{-1}F_{B1}F_{AB}F_{B1}^{-1}$. According to the nontrivial fusion rule of MZMs, two MZMs would fuse either as a vacuum state $I$ or an unpaired fermionic state $\Psi$. The system's state thus shifts to a superposition of $|0_{\text{AB}}0_{12} \rangle $ and $|1_{\text{AB}}1_{12} \rangle $, implying a half charge transferring from TS to FM. As revealed by the yellow dashed line in Fig. \ref{f2}(a), a half charge indeed transfers from TS to FM after a complete loop. However, a FM state is usually not stable at the zero energy. If $E_d$ remains positive, the splitting process would not happen and the system simply returns to $|0_{\text{AB}}0_{12} \rangle $ state after one period. In this case, there's no charge transfer as revealed by the red solid line shown in Fig. \ref{f2}(a), even though the deviation is small (the corresponding $E_d(t)$ is shown by the red solid line in Fig. \ref{f2}(c)). If instead the energy of the FM crosses zero, say for $a = 0.9$, the FM state would switch to the occupied state $|1_{\text{AB}} \rangle $, and the whole system would choose $|1_{\text{AB}}1_{12} \rangle $ finally due to the conservation of fermion parity. This results in an integer charge transferring from FM to MZMs after one complete loop as  red solid line in Fig. \ref{f2}(a) shows. Such a nontrivial pumping cannot happen in the trivial loop denoted by $F_{AB}^{-1}F_{B1}F_{B1}^{-1}F_{AB}$, where the net charge transfer is always zero as indicated by Fig. \ref{f2}(b), regardless whether $E_d$ crosses zero level during the whole process as can be seen in Fig. \ref{f2}(d).

Since the splitting of MZMs $\gamma_A$ and $\gamma_B$ takes place at $E_d=0$, the system parameters at this moment play a pivotal role in determining the occurrence of nontrivial pumping. As indicated by path 1 in Fig. \ref{f3}(c), we set $E_d(t) =+(-) E_0\cos(\omega t)$ for  $t\in[0,T](t\in [2T,3T])$. In this situation, the final fusion energy of MZM transitions from $2E_0$ to $E_0$, while $E_d(t)$ and $t_c(t)$ both stay at zero energy at $t=T/2$ (or $\omega t = \pi/2)$ and the parameters cross one time as indicated by the circle.  Consequently, an integer charge is transferred as revealed by the red solid line in Fig. \ref{f3}(a), which is independent of the final fusion energy. However, if we fuse  $\gamma_B$ and $\gamma_1$ before $E_d(t)$ reaches zero (path 2 in Fig. \ref{f3}(c)), the net charge transfer would be zero after the loop is completed. This suggests that the nontrivial charge transfer directly relate to the crossing between $E_d$ and $t_c$ at zero energy in the parameter space. Intriguingly, path 3 mirrors path 1 but with an extra intersection of $E_d(t)$ and $t_c(t)$ at zero energy (shown by the red dashed line in Fig. \ref{f3}(d)). This results in the charge transfer reverting to zero. Conversely, by varying $t_c(t)$ as depicted in path 4 of Fig. \ref{f3}(d), where only a single zero-energy crossing occurs, the final charge transfer would remain to be one, as seen in Fig. \ref{f3}(c). These results suggest a simple criterion for nontrivial charge pumping: if $E_d(t)$ and $t_c(t)$ intersect at zero energy an odd number of times in a closed loop, the charge pumping becomes nontrivial. Conversely, an even number of such intersections within the interval results in trivial charge pumping. This observation holds potential to guide experimental manipulations. Practically, tuning $t_c$ to zero may pose a primary challenge in our proposal. However, we expect the nontrivial pumping to be observed when the time period $T \ll 1/ \min(t_c)$, where $\min(t_c)$ represents the minimum value of $t_c$ at the moment of intersection. Note that $t_c$ is determined by the tunneling barrier and usually decays exponentially with barrier size, which can be modulated by gate voltage. Adiabatic evolution further requires $T \gg 1/\Delta$. As a result, the time window for the nontrivial pumping is $1/\Delta \ll T \ll 1/ \min(t_c)$.




\begin{figure}
  \centering
  \includegraphics[width=3.25in]{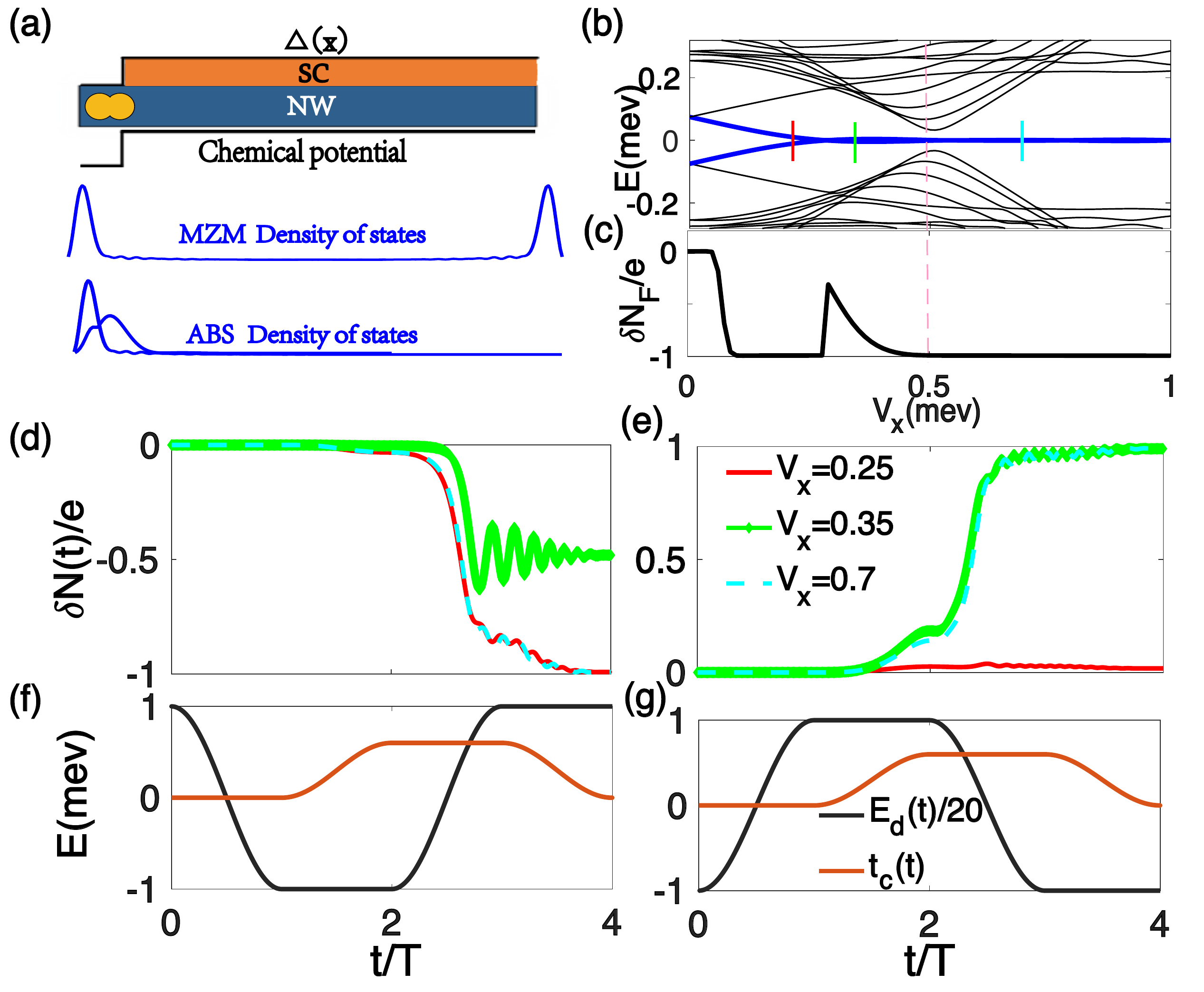}
  \caption{
  (a) Schematic plot of unpaired MZMs (upper) and ABS (lower) in a nanowire. We use a hard wall confinement to introduce ABS (a step-like potential depicted as the black  line).
  (b) Evolution of energy spectrum with external magnetic field in the NS system. The red dashed line indicates the topological phase transition points. On the left side, the system is in the trivial regime, with ABS localized at the interface.
  (c) The variation of charge pumping with external magnetic field in the NS system according the parameters of (f).
  (d) The charge transfer process for $V_x=0.7 \text{ mev}$ (cyan cut in (b)), $V_x=0.35 \text{ mev}$ (green cut in (b)), and $V_x=0.25 \text{ mev}$ (red cut in (b)), with $E_d$ and $t_c$ varying according to (f).
  (e) The charge pumping versus time $t$ for three representative cases as in (d). The variations of the parameters $E_d$ and $t_c$ with time are depicted in (g). Here FM is occupied, and therefore the charge pumping is from FM to TS.
  }
  \label{f4}
\end{figure}

\textit{Numerical simulation in the nanowire superconducting nanowire.}  We have shown that the nontrivial charge pumping would be induced through controlling the order of both fusion and splitting processes among the MZMs. Naturally, one may wonder how ABS and MZM differ during this process. To investigate this, we further simulate the non-Abelian fusion processes in a realistic nanowire/superconductor (NS) system as depicted in Fig. \ref{f4}(a). In general, the ABS state may arise due to inhomogeneous potential or disorder at the interface of the NS system \cite{S3}.  Here, we introduce a typical hard wall confinement at the end of the NS to model the inhomogeneous potential. This results in a nearly zero-energy ABS state being trapped at the left end of NS before the system is driven into topologically nontrivial phases by increasing Zeeman energy $V_x$, as shown in Fig. \ref{f4}(b). In practice, the chemical potential of the FM can be modulated through gate voltage as demonstrated in recent experiment, so is the coupling between the FM and the NS system \cite{Pgate1,Pgate2}. Thus our fusion protocol could be readily implemented in real systems.


In Fig. \ref{f4}(c), we show the final charge transfer from FM to TS for the nontrivial loop depicted in Fig. \ref{f4}(f) when the Zeeman energy $V_x$ varies, which suggests that ABS states can also introduce nontrivial charge pumping. Representative charge transfer processes are demonstrated in Fig. \ref{f4}(d) for $V_x =0.25$ (red line), $0.35$ (blue line) and $0.7$ (cyan line). Clearly, an integer charge is transferred when $V_x =0.25$ and $V_x =0.7$. For $V_x =0.35$, however, only a fractional charge is transferred after a complete loop. We may also set $E_0=-0.05 \text{ mev}$ at $t=0$ as shown in Fig. \ref{f4} (g), where the FM is initially occupied. The resulting charge transfer  as shown in Fig. \ref{f4}(e) is solely determined by the fusion channel of the FM. If the fermionic state is initially unoccupied (corresponding to fusion channel $I$), the charge should be transferred from MZM to FM. This transfer process will be reversed if the fermionic state is initially occupied (fusion channel $\Psi$). Moreover, integer pumping is only robust in the topological region,  while the quantized charge pumping for $V_x =0.25$ is failed for the occupied case and is quantized for $V_x=0.35$. This means that the quantized charge pumping is robust for MZMs, regardless whether the initial FM state is occupied or unoccupied. As for ABS,   it can be divided into two kinds of states. One is the electron-like ABS which can support quantized charge transfer if the initial FM state is occupied. Another one is the hole-like ABS which can support quantized charge transfer if the initial FM state is unoccupied.


The distinct behaviors of MZMs and ABS can be further understood from transport processes. During the fusion,  the charge transfers through either TE or TH process, not both (see Supplementary materials for more information), determined by the total parity of the system. To understand this, we first note that the total parity is conserved with the evolution of $E_d$ and $t_c$. If the initial state is $|0_{\text{AB}}0_{12} \rangle $ which means both the FM and MZM are unoccupied, it would gradually switch to state $|1_{\text{AB}}1_{12}\rangle$ after half a period. The TE process, which only transfers electrons between FM and MZM, cannot make it. Hence, the pumping has to be accomplished through TH process, during which a hole in the FM is transformed to an electron in MZM. However, if $E_0<0$, the initial state would be $|1_{\text{AB}}0_{12}\rangle$ instead and only TE process occurs. Since MZM is its own anti-particle, there's no difference between TE and TH process. While for ABS one of the two processes would be preferred as it is not hermitian.

\textit{Conclusion.} With the assistance of a fermionic mode, we unveil the fusion rule of MZMs, which manifests itself in the nontrivial charge transfer process. The FM can be easily attached either to the end of 1D TS wire or a vortex in 2D TS system. Moreover, by controlling the order of fusion and splitting processes among MZMs, one may design trivial and nontrivial fusion loops tailored for the convenience of specific platforms since each process can be individually adjusted. Therefore, our proposal can be applied to various TS systems without additional constraints. In our proposal, one only needs to tune the energy level of FM as well as its coupling to the Majorana modes through gate voltages, leaving the parameters of topological superconductor intact. 
This may avoid potential complications arising from meddling with TS wires. Finally, we emphasize that the charge transfer are distinct for different situation,  we can expect that other realistic factors, such as finite temperature or interaction, only play insignificant roles in our setup. Thus, our proposal provides a feasible route for showcasing the nontrivial fusion rules of MZMs.

\textit{Acknowledgement.} This work is financially supported by National Natural Science Foundation of China (Grants No. 11974271, and No. 92265103), the National Basic Research Program of China (Grants No. 2015CB921102 and No. 2019YFA0308403), and the Innovation Program for Quantum Science and Technology (Grant No. 2021ZD0302400).

\clearpage
\onecolumngrid

\setcounter{equation}{0}
\setcounter{figure}{0}
\setcounter{page}{1}

\begin{center}
    {\textbf{\large Supplemental Material for ``Unveiling non-trivial fusion rule of Majorana zero mode using a fermionic mode"}\\[1cm]}
\end{center}

\appendix








\renewcommand{\theequation}{S\arabic{equation}}
\renewcommand{\thefigure}{S\arabic{figure}}
\renewcommand{\bibnumfmt}[1]{[S#1]}
\renewcommand{\citenumfont}[1]{S#1}

\section{Additional nontrivial fusion loop}
\begin{figure}[b]
\centering
\includegraphics[width=\textwidth]{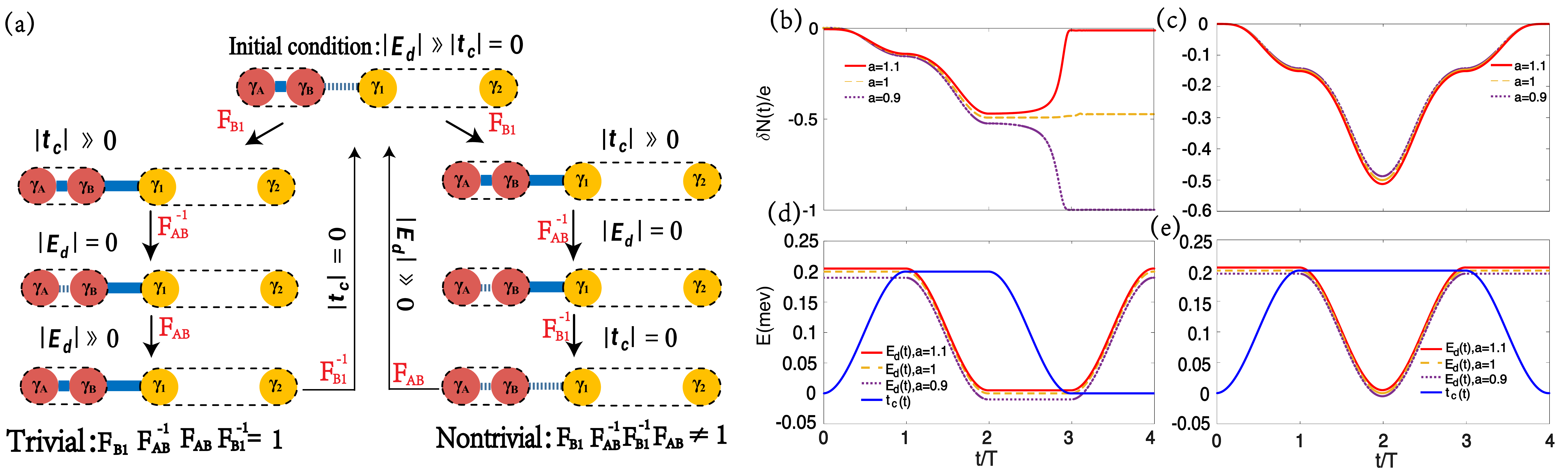}
\caption{
(a) Modified loops compared to the ones shown in the main text. In the trivial loop (left panel), we first fuse $\gamma_B$ and $\gamma_1$ ($F_{B1}$), followed by the splitting of $\gamma_A$ and $\gamma_B$ ($F_{AB}^{-1}$), and then reverse the operation to complete the loop. In the nontrivial loop (right panel), two consecutive processes do not commute with each other.
(b)  The charge transfer for the nontrivial loop $F_{B1}F_{AB}^{-1}F_{AB}F_{B1}^{-1}$. Here $F_{AB}^{-1}$ ($F_{AB}$) is controlled by $E_d(t) = E_0[a\pm \cos(\omega t)]$ with $\omega=1/1000 \text{ mev}$ and $E_M=10^{-9} \text{ mev}$. After the splitting process, the energy level of the FM reaches exactly to zero for $a = 1$, stays above zero for $a = 1.1$, and changes sign for $a=0.9$.
(c) The charge transfer for the the trivial loop $F_{B1}F_{AB}^{-1}F_{AB}F_{B1}^{-1}$.
(d) and (e) show the evolution of system parameters corresponding to (b) and (c), respectively.}
\label{R1}
\end{figure}

As we have shown in the main text,  the trivial and nontrivial fusion loops can be designed by manipulating the order of fusion and splitting processes.  
Contrary to the fusion loop in the main text, here we fuse $\gamma_B$ and $\gamma_1$ first. As such, the trivial loop  would be  denoted as $F_{B1}F_{AB}^{-1}F_{AB}F_{B1}^{-1}$ and the nontrivial loop could be denoted as $F_{B1}F_{AB}^{-1}F_{B1}^{-1}F_{AB}$, as shown in Fig. \ref{R1}(a). The charge transfer for the nontrivial loop shows a zero, half, integer charge pumping in 
Fig. \ref{R1}(b) when the corresponding parameters vary as Fig. \ref{R1}(d) depicts.  While for the trivial case the net charge transfer is always zero as indicated by Fig. \ref{R1}(c), regardless whether $E_d$ crosses zero level during the whole process, as can be seen in Fig. \ref{R1}(e). Although it shows the same charge transfer behavior as the charge pumping in Fig. 2 of the main text, the charge pumping here is more stable and does not oscillate in the case $E_d=0$ with $a=1$. 
This is because the charge would have no definite way to transfer in and out of the FM states at $E_d=0$ and display an oscillating behavior when we vary $t_c$ during the loops shown in the main text. Nevertheless the modified loop here can avoid such uncertainty.

\begin{figure}
\centering
\includegraphics[width=5.25in]{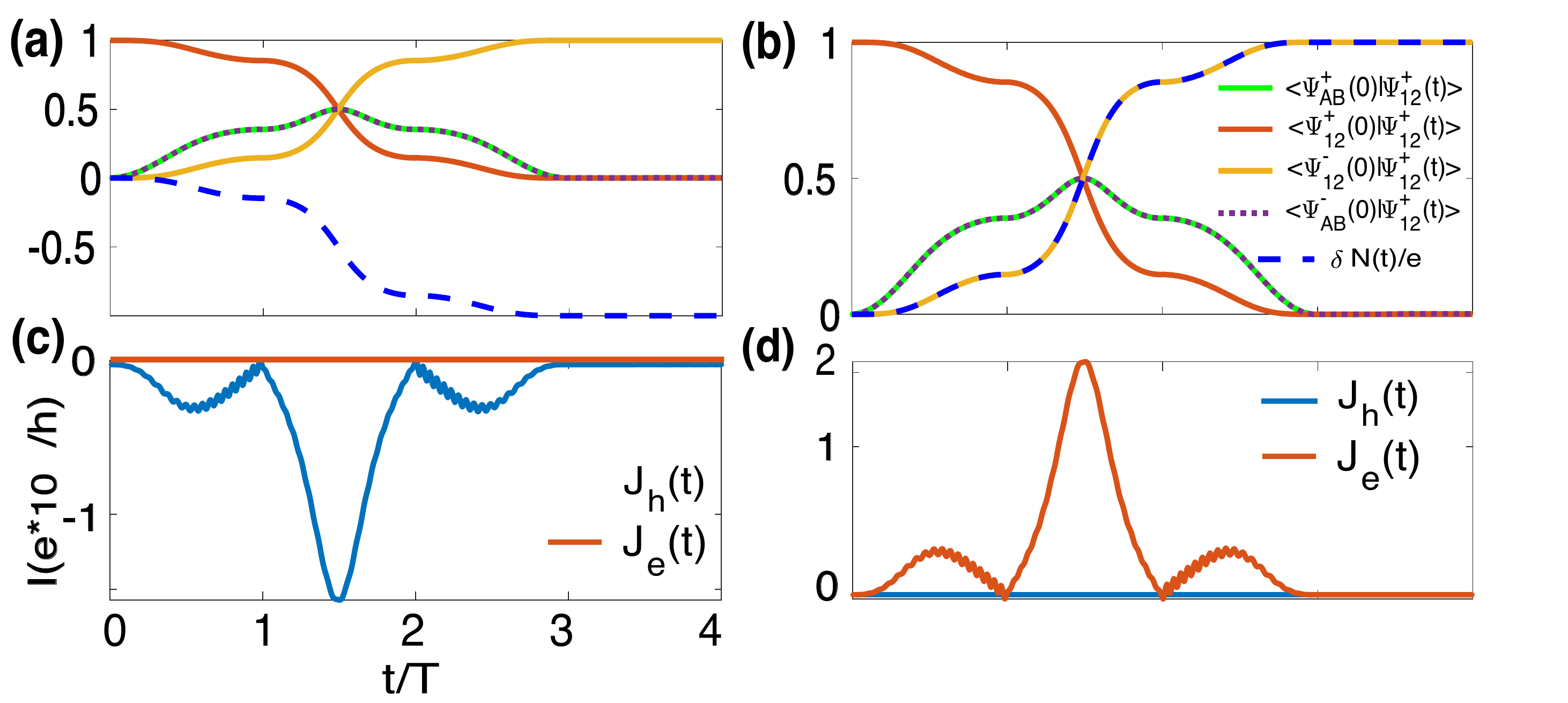}
\caption{ The charge transfer and the evolution of the wavefunction $\psi_{B1}^{+}(t)$ following the loop  $F_{B1}F_{AB}^{-1}F_{AB}F_{B1}^{-1}$ with the initial state  $E_d(0)>0$ in (a) and  $E_d(0)<0$ in (b). 
Here  the fusion ($F_{AB}$) and splitting process ($F_{AB}^{-1}$) is controlled by $E_d(t) = E_0[a\pm \cos(\omega t)]$ with $a=0$. $E_0$ is a positive value in (a) and a negative value in (b). The other parameters are the same as Fig. \ref{R1}. Although the evolution of wavefunction are the same in both situations, they choose different transfer process. The transfer process  in (a)  is TH process as shown in (c)  since the parity of initial state is even for $E_d(0)>0$, while the transfer process  in (b)  is TE process as shown in (d) since the parity of initial state is odd for $E_d(0)<0$.}
\label{R2}
\end{figure}

Moreover,  the initial conditional for $E_d>0$ and $E_d<0$ is very important for the whole fusion process. Since we set the initial state composed by MZMs $\gamma_1$ and $\gamma_2$ as $|0_{12}\rangle$, the parity of the whole system would
be completely determined by the initial sign of $E_d$. If $E_d>0$, the initial state is $|0_{\text{AB}}0_{12} \rangle $ which means both the FM and MZMs are unoccupied, and the system is in even-parity state during the loop. If $E_d<0$, however, the initial state would be $|1_{\text{AB}}0_{12} \rangle $ which means 
FM is occupied and MZM is unoccupied with the total parity being odd. Figure \ref{R2}(a) reveals the evolution of wavefunction $\psi_{12}^ {+}(t)$ in the nontrivial fusion loop. Here $\psi_{12}^ {+}(t)$ is the evolution state of $\psi_{12}^ {+}(0)$ at time $t$, and $\psi_{AB}^{\pm}(0)=\gamma_A\pm i\gamma_B$, $\psi_{12}^{\pm}(0)=\gamma_1 \pm i\gamma_2$. As we can see, $\psi_{12}^ {+}(t)$ would transform as $\psi_{12}^{-}(0)$ after the whole fusion loop. This means the initial state is $|0_{\text{AB}}0_{12} \rangle $,  and it would gradually switch to state $|1_{\text{AB}}1_{12}\rangle$ after a complete loop. During the evolution, an integer charge would be pumped to FM.  The pumping is
accomplished through TH process. This is consistent with the current plot in Fig. \ref{R2}(c). In the whole process, $J_e(t)$, the current induced by TE process, remains to be zero, while $J_h(t)$, the current induced by TH process, takes finite value. 
Figure \ref{R2}(b) reveals the evolution of wavefunction $\psi_{12}^ {+}(t)$ in the situation $E_d<0$, where the initial state is $|1_{\text{AB}}0_{12} \rangle $,  and gradually switches to state $|0_{\text{AB}}1_{12}\rangle$ in the end of the loop. The corresponding pumping is accomplished through TE process as shown in Fig. \ref{R2}(d).

\section{Reduced step for nontrivial fusion}
\begin{figure}
\centering
\includegraphics[width=5.25in]{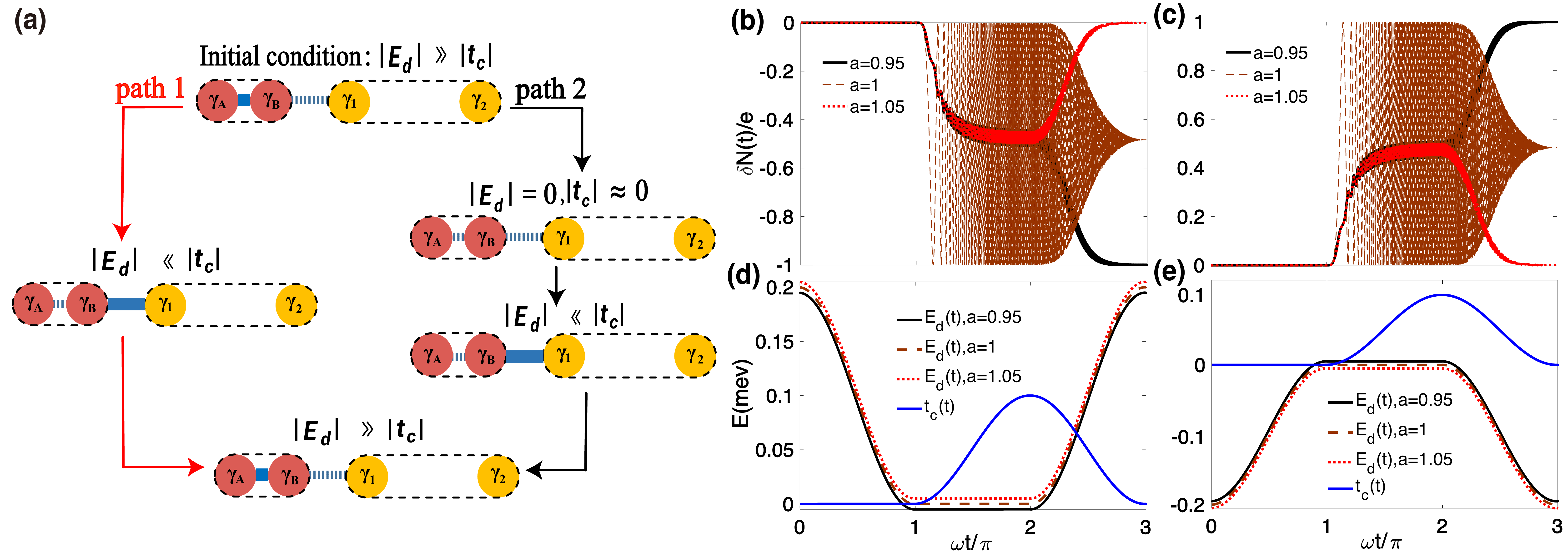}
\caption{(a)  Path 1 represents the trivial fusion loop with only two steps and path 2 is the non-trivial fusion loop with three steps. For path 2, we set the parameters as $E_d(t) = E_0[a+\cos(\omega t)]$ for $\omega t\in[0,\pi]$, then $t_c(t)=|E_0|[1+\cos(\omega t)]/2$ for $\omega t\in[\pi,2\pi]$, and finally $E_d(t) = E_0[a-\cos(\omega t)]$ and $t_c(t)=|E_0|[1+\cos(\omega t)]/2$ for  $\omega t\in[2\pi,3\pi]$. 
(b) Charge transfer at $a=1.05$, $1$, and $0.95$ with $E_0=0.1\text{ ev}$. Since the charge chooses different fusion channel for $a>1$ and $a<1$, the final charge transfer is different (zero for $a>1$ and one for $a<1$).
(c) The charge transfer for  $E_0=-0.1\text{ ev}$. The current following direction is reverse. (d) and (e) show the correspond manipulation of $E_d$ and $t_c$  in (b) and (c), respectively.}
\label{R3}
\end{figure}

Since the charge pumping is determined by the number of intersections of $E_d(t)$ and $t_c(t)$ at zero energy, we can then control the pumping charge by adjusting both the fusion and splitting steps, as illustrated in Fig. \ref{R3}(a). For the trivial loop, $E_d$ and $t_c$  can be  modulated at the same time as $E_d = E_0\cos(\frac{\omega t+\phi_0}{2})$, $t_c=|E_0|[1-\cos(\omega t+\phi_0)]/2$. 
In this situation,  the fusion loop can be reduced to two steps. For the nontrivial case as shown in Fig. \ref{R3}(b), the first two steps of $F_{AB}^{-1}$ and $F_{B1}$ are gated individually and the other two steps can be adjusted at the same time. To be specific,  we set $t_c=0$ and $E_d(t) = E_0[a+\cos(\omega t)]$ in the first step with $\omega t\in[0,\pi]$, then $t_c(t)=|E_0|[1+\cos(\omega t)]/2$ for $\omega t\in[\pi,2\pi]$, and finally the splitting and fusion processes can be realized at the same time by setting $E_d(t) = E_0[a-\cos(\omega t)]$ and $t_c(t)=|E_0|[1+\cos(\omega t)]/2$ for $\omega t\in[2\pi,3\pi]$.  Clearly, the charge transfer displays the same behavior as when we manipulate each step individually in the main text.  In the case of $a=1$, which means $E_d$ is tuned exactly to zero energy after the first step as shown by the dashed line in Fig. \ref{R3}(d), the system would have no definite way to evolve in the subsequent step. As we increase $t_c$ to be $t_c\gg E_d\approx 0$, the system would be in a superposition state of $|0_{\text{AB}}0_{12} \rangle $ and $|1_{\text{AB}}1_{12} \rangle $. After the couplings recover the initial values through the final step, such a nontrivial fusion process would introduce a half charge transfer, as revealed by the dashed line in Fig. \ref{R3}(b).
A small deviation with $a = 1.05$ would result in zero charge transfer as revealed by the red dotted line shown in Fig. \ref{R3}(b) (The corresponding $E_d(t)$ is shown by the red dotted line in Fig. \ref{R3}(d)). However, if $a = 0.95$, the energy of FM would change sign and the system chooses $|1_{\text{AB}}1_{12} \rangle $, resulting in an integer charge transferred from FM to MZMs after one period as black solid line in Fig. \ref{R3}(b) shows. 

\begin{figure}
\centering
\includegraphics[width=5.25in]{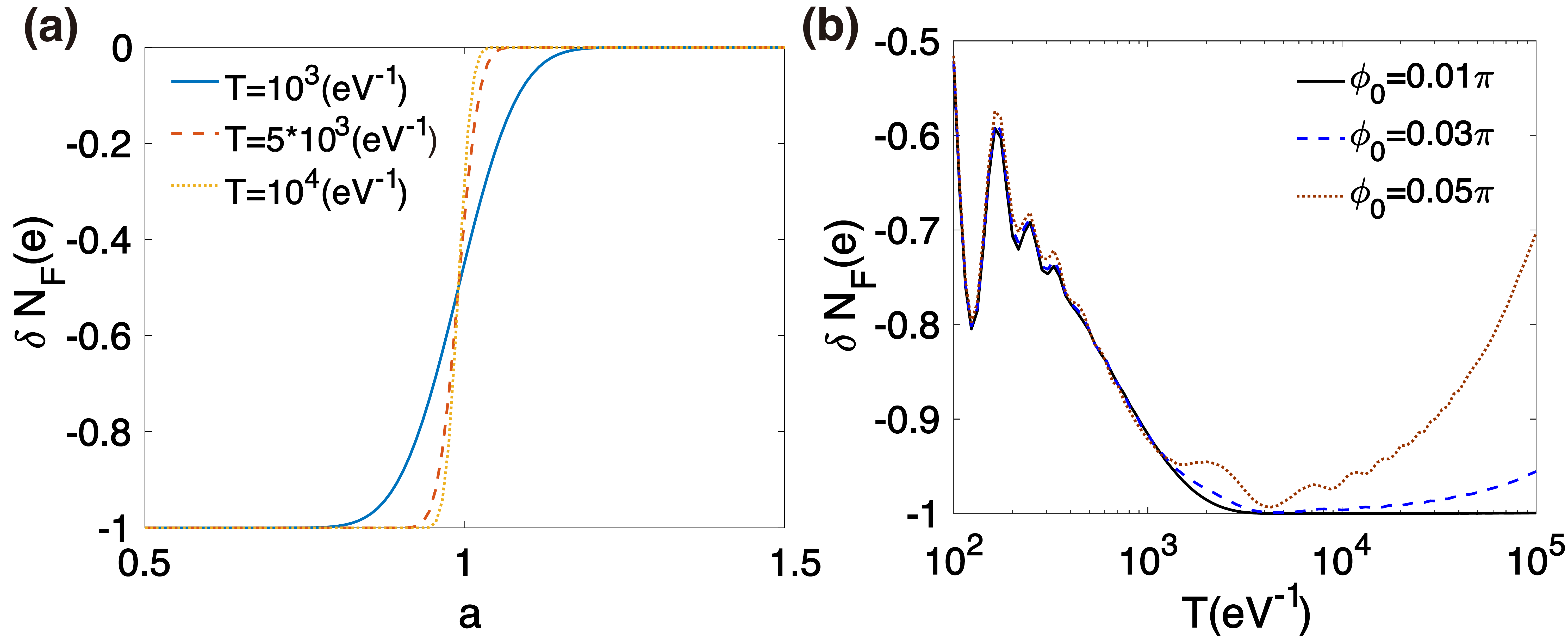}
\caption{ (a) The charge transfer is distinct for $a>1$ and $a<1$ with different time period. There's an integer charge transfer if $a<1$, and no charge transfer if $a>1$.  Since such process is an adiabatic process,  the transition region would become more and more steep with time increase. (b) The influence of final charge transfer in the case of $t_c \neq 0$. We set $t_c(t)=|E_0|(1-\cos(\phi_0))/2$ in the initial time. 
This means  $t_c \neq 0$ for $\phi_0\neq 0$, It would gradually destroy nontrivial charge transfer process but still sustains in a finite time window}
\label{R4}
\end{figure}

The charge transfer would be more stable with a slowly period. As further revealed in Fig. \ref{R4}(a), for an initial state $|0_{\text{AB}}0_{12} \rangle $, $E_d$ would cross zero for $a<1$ after the first step, then the states would transfer to $|1_{\text{AB}}1_{12} \rangle $ state, which would cause an integer charge pumping from MZM to FM. If $a>1$, $E_d$ would remain above zero energy all the time, then the states would transfer back to $|0_{\text{AB}}0_{12} \rangle $, with no net charge transfer in the end. We investigate the charge transfer process for different period $T=10^{4} (eV)^{-1}$, $T=0.5*10^{4}(eV)^{-1}$, $T=10^{3}(eV)^{-1}$ with $\omega T=\pi$. We can see that the  charge pumping $\delta N_F$ after the manipulation is definite with slowly oscillation period.
Since we keep $t_c$ at zero in the first step as $E_d$ varies, we further study the case for $t_c \neq 0$. We investigate the parameters varying as $E_d(t) = E_0[a+\cos(\omega t)]$ and $t_c(t)=|E_0|[1-\cos(\phi_0)]/2$ for $\omega t\in[0,\pi]$, then $t_c(t)=|E_0|[1-\cos(\omega t+\phi_0)]/2$ for $\omega t\in[\pi,2\pi]$, and finally $E_d(t) = E_0[a-\cos(\omega t)]$ and $t_c(t)=|E_0|[1+\cos(\omega t+\phi_0)]/2$ for  $\omega t\in[2\pi,3\pi]$.  It is the same as the parameters of Fig. 4 in the main text except that there's an initial deviation $\phi_0$ in $t_c$. We show the final charge transfer versus period $T$ at $a=0.9$ in Fig. \ref{R4}(b). 
We find that the nontrivial charge transfer can still be distinguished even for $\phi_0=0.05\pi$ for $T\in[10^{3},10^{4}](eV)^{-1}$. This suggests that the nontrivial pumping to be observed when the time period $T \ll 1/ \min(t_c)$, where $\min(t_c)$ represents the minimum value of $t_c$ at the moment of intersection. Moreover, adiabatic evolution further requires $T \gg 1/\Delta$. As a result, the time window for the nontrivial pumping is $1/\Delta \ll T \ll 1/ \min(t_c)$.

\section{The Hamiltonian and parameters for tight-binding model}

The tight binding Hamiltonian of the NS system is given by
\begin{eqnarray}
\label{model}
 H_{NS} &=& \sum\nolimits_{\mathbf{R},\mathbf{d},\alpha} { - t_0(\psi _{\mathbf{R} + \mathbf{d},\alpha }^\dag \psi_{\mathbf{R},\alpha } + h.c.) - \mu \psi_{\mathbf{R},\alpha }^\dag \psi_{\mathbf{R},\alpha } } \nonumber \\
&+& \sum\nolimits_{\mathbf{R},\mathbf{d},\alpha ,\beta } { - i{U _R} \psi _{\mathbf{R} + \mathbf{d},\alpha }^\dag  \hat z \cdot (\vec{\sigma}  \times \mathbf{d})_{\alpha \beta }   \psi _{\mathbf{R},\beta } } \nonumber \\
&+& \sum\nolimits_{\mathbf{R},\alpha} \Delta(R)e^{i\phi} \psi _{\mathbf{R},\alpha }^{\dagger} \psi _{\mathbf{R},-\alpha }^{\dagger} + h.c. \nonumber \\
&+& \sum\nolimits_{\mathbf{R},\alpha ,\beta } { \psi _{\mathbf{R}, \alpha }^\dag (V_x\vec{\sigma}_x)_{\alpha \beta} \psi _{\mathbf{R}, \beta}}.
\end{eqnarray}

\noindent Here, the subscript  $\mathbf{R}$ denotes the lattice site; $\mathbf{d}$ is the unit vector and $\mathbf{d}_{x}$ connects the nearest neighbor sites along $x$- direction; $\alpha$ and $\beta$ are the spin indices; $t_0$ denotes the hopping amplitude; $\mu$ is the chemical potential; $U_{R}$ is the Rashba coupling strength; and $V_x$ is the Zeeman energy. The superconducting pairing amplitude is denoted as $\Delta$, and $\phi$ is the pairing phase. In the numerical simulations, the effective mass  $m^{*}=0.026m_e$, the Rashba spin-orbit coupling strength $\alpha=30\text{ mev}\cdot \text{nm}$, the superconducting pairing strength $\Delta=0.25 \text{ mev}$, the g factor $g=15$, the length of nanowire is $N_x=120a$ with he lattice constant $a = 20 \text{ nm}$,  The quantum dot which connects with the nanowire is described by
\begin{equation}
 	 H_d=\sum\nolimits_{\alpha}2E_dd^\dagger_\alpha d_\alpha+\sum\nolimits_{\alpha ,\beta } { d _{ \alpha }^\dag (V_x\vec{\sigma}_x)_{\alpha \beta} d _{ \beta}},
\end{equation}

\noindent where $E_d$ represents the on-site energy of the QD and $d_\alpha$ is the fermionic annihilation operator with spin $\alpha$. The coupling between the nanowire and the QD is described by the hopping term

\begin{equation}
	H_{Sd}=\sum\nolimits_{\alpha}t_c\psi^\dagger_{1,N_x,\alpha}d_{\alpha}+h.c.,
\end{equation}

\noindent where $t_c$ is the coupling strength. 

In experiment, the trivial Andreev bound states (ABSs) could appear at zero energy and blend with MZMs. It is necessary to investigate the fusion rule of trivial ABSs.  Such ABS is usually induced by a hard-wall confined potential at the interface, which can be modeled by a square quantum well confinement at the end of the nanowire \cite{Marcus, Aguado, ChunXiao1}.
We simulate the square quantum well by setting $\mu(R) = -2t$, $\Delta(R) = 0$ for  $\mathbf{R}<20a$ and $\mu(R) = -2t+0.45\text{ mev}$ , $\Delta(R) = \Delta$ for $\mathbf{R}>20a$.  When an external magnetic field $V_{x1}$ and a smooth confinement is applied in the nanowire, as shown in Fig. 4(b) of main text, ABS will be trapped in the confinement region before the topological phase transition occurs ($0.2\text{ mev}<V_{x1}<0.45 \text{ mev}$). In contrast to the MZMs that distribute non-locally at both ends of the nanowire, the quasi-MZM can be viewed as two strongly overlapped MZMs $\gamma_1$ and $\gamma_2$ which are both localized inside the confinement region. To numerically simulate the fusion process in a NS wire, we follow the traditional time-evolution method of trotter decomposition \cite{Lei}.

\section{Exact solution of FM-MZM hybrid system}
In the main text we have shown a simple criterion for nontrivial charge pumping: if $E_d(t)$ and $t_c(t)$ intersect at zero energy an odd number of times, the charge pumping becomes nontrivial, and the opposite is true for even intersections. 
Such rule is different from  Thouless pumping. In Thouless pumping, the charge pumping is determined by the winding number encircled by the parameters. However, we can still find a exact solution for this system.
We start with the Hamiltonian $H_M$ in the Majorana representation, given by:

\begin{equation}
\begin{aligned}
H_M &= i{E_d}\gamma_A\gamma_B + i|t_c|\gamma_B{\gamma_1}.
\end{aligned}
\label{eq2}
\end{equation}
In the qubit operator representation: $\sigma_z=i\gamma_A\gamma_B$ and $\sigma_x=i\gamma_B\gamma_1$, and the Hamiltonian exactly describes a typical two-level system. Since $\sigma_z$ and $\sigma_x$ do not commute with each other, tuning $E_d$ and $t_c$ in different order would lead to distinct results.  However, since the charge pumping only depends on the intersections of system parameters during the evolution in a complete loop, we do not have to care about the specific forms of the paths such as the form of $E_d(t)$ or $t_c(t)$. The nontrivial loop can thus be homotopy to a loop with two periods: 
in the first period, we vary
$E_d$ from $E_0$ to $-E_0$ while $t_c$ keeps zero.  In the second part, we set $E_d=-E_0 \cos(\omega t)$ and $t_c =E_0\sin(\omega t)$ from $\omega t=0$ to $\omega t=\pi$. This is just a half circle in the parameter space. For the first part, the evolution would induce a dynamical phase $\phi = \int_0^TE_d(t)\sigma_zdt$.  While for the second part, it is just a nuclear magnetic resonance system with $H_N(t) = -E_0 \cos(\omega t)\sigma_z+E_0 \sin(\omega t)\sigma_x$. This Hamiltonian can be exactly calculated through unitary transformation $\hat{U}(t) =\exp(-i\frac{\omega t}{2}\sigma_y)$. Then the effective Hamiltonian would become time independent:
\begin{equation}
\begin{aligned}
H_{eff} =\hat{U}(t)H_N(t)\hat{U}(t)^{-1}- \hat{U}(t)(-i\frac{\partial}{\partial t}\hat{U}(t)^{-1})= -E_0\sigma_z-\frac{\omega}{2}\sigma_y.
\end{aligned}
\label{eq2}
\end{equation}

Now the time evolution operator $U$ can be given by:
\begin{equation}
\begin{aligned}
U(t) =\hat{U}(t)e^{-iH_{eff}t}= \exp(-i\frac{\omega t}{2}\sigma_y)[\cos(\Omega t)-i\sin(\Omega t)(\frac{E_0}{\Omega}\sigma_z-\frac{\omega}{2\Omega}\sigma_y)]
\end{aligned}
\label{eq2}
\end{equation}

Here, $\Omega=\sqrt{E_0^2+(\omega/2)^2}$. If $\omega\ll E_0$, then $\Omega\simeq E_0$ and $U(t)\simeq \exp(-i\frac{\omega t}{2}\sigma_y)\exp(iE_0t\sigma_z)$. In this situation, the states would be slowly rotated with frequency $\omega/2$. Then the state would change from $|0_{\text{AB}}0_{12} \rangle $ to $|1_{\text{AB}}1_{12} \rangle $ in the end of the loop.


\begin{thebibliography}{WSM_JJ}
\bibitem{S1} We should note that, $E_M$ is not strictly zero in realistic setup and usually takes a very small value.  Here, we set $E_M$ as a positive value, so the initial parity of MZM is $|0_{12} \rangle $. If, however a trivial ABS instead of MZM appears at the end of nanowire, we could simply set $E_M$ to be a finite value comparable with $E_d$ or $t_c$.
\end{thebibliography}
\end{document}